\documentclass[aip,reprint,superscriptaddress,nofootinbib,twocolumn,10pt]{revtex4-2}

\usepackage[linktoc=page,colorlinks,urlcolor=blue,citecolor=blue,linkcolor=blue]{hyperref}
\usepackage{amssymb,amsmath}
\usepackage{graphicx}
\usepackage{natbib}
\usepackage{lineno}
\usepackage[dvipsnames]{xcolor}
\usepackage{overpic}
\usepackage[font=small,labelfont=bf,
   justification=raggedright,
   format=plain]{caption}
\newcommand{\snl}{Sandia National Laboratories, Albuquerque, New Mexico 87185, USA}
\newcommand{\cint}{Center for Integrated Nanotechnologies, Sandia National Laboratories, Albuquerque, New Mexico 87123, USA}
\newcommand{\unm}{University of New Mexico Department of Physics and Astronomy, Albuquerque, New Mexico 87131, USA}
\newcommand{\brown}{Department of Physics, Brown University, Providence, Rhode Island 02912, USA}
\newcommand{\NIMS}{National Institute for Materials Science, 1-1 Namiki, Tsukuba 305-0044, Japan}
\newcommand{\hydrogen}{$^1$H}
\newcommand{\boron}{$^{11}$B}
\newcommand{\carbon}{$^{13}$C}
\newcommand{\nitrogen}{$^{15}$N}
\newcommand{\fluorine}{$^{19}$F}

\usepackage{verbatim}

\begin{document}

\title{Nanoscale Solid-State Nuclear Quadrupole Resonance Spectroscopy using Depth-Optimized Nitrogen-Vacancy Ensembles in Diamond}

\date{\today}
\author{Jacob Henshaw}\affiliation{\cint}
\author{Pauli Kehayias}\affiliation{\snl}
\author{Maziar Saleh Ziabari}\affiliation{\cint}\affiliation{\unm}
\author{Michael Titze}\affiliation{\snl}
\author{Erin Morissette}\affiliation{\brown}
\author{Kenji Watanabe}\affiliation{\NIMS}
\author{Takashi Taniguchi}\affiliation{\NIMS}
\author{J.I.A. Li}\affiliation{\brown}
\author{Victor M. Acosta}\affiliation{\unm}
\author{Edward Bielejec}\affiliation{\snl}
\author{Michael P. Lilly}\affiliation{\cint}
\author{Andrew M. Mounce}\affiliation{\cint}
\begin{abstract}
Nuclear magnetic resonance (NMR) and nuclear quadrupole resonance (NQR) spectroscopy of bulk quantum materials have provided insight into phenomena such as quantum phase criticality, magnetism, and superconductivity. With the emergence of nanoscale 2-D materials with magnetic phenomena, inductively-detected NMR and NQR spectroscopy are not sensitive enough to detect the smaller number of spins in nanomaterials.  The nitrogen-vacancy (NV) center in diamond has shown promise in bringing the analytic power of NMR and NQR spectroscopy to the nanoscale. However, due to depth-dependent formation efficiency of the defect centers, noise from surface spins, band bending effects, and the depth dependence of the nuclear magnetic field, there is ambiguity regarding the ideal NV depth for surface NMR of statistically-polarized spins. In this work, we prepared a range of shallow NV ensemble layer depths and determined the ideal NV depth by performing NMR spectroscopy on statistically-polarized \fluorine{} in Fomblin oil on the diamond surface. We found that the measurement time needed to achieve an SNR of 3 using XY8-N noise spectroscopy has a minimum at an NV depth of 5.4 nm. To demonstrate the sensing capabilities of NV ensembles, we perform NQR spectroscopy on the \boron{} of hexagonal boron nitride flakes. We compare our best diamond to previous work with a single NV and find that this ensemble provides a shorter measurement time with excitation diameters as small as 4 $\mu$m. This analysis provides ideal conditions for further experiments involving NMR/NQR spectroscopy of 2-D materials with magnetic properties.
\end{abstract}
\maketitle
\section{Introduction}

Nuclear magnetic resonance (NMR) and nuclear quadrupole resonance (NQR) spectroscopy have been useful tools for interrogating magnetic ordering and dynamics in materials for over half a century \cite{NMRSC-slichter}. When used to probe nuclei in solids, these tools provide an atomic-scale perspective of the electronic and magnetic proprieties of these materials\cite{AFM_LowDmag,FM_SmMnGe,CurroHF_critical,NMRSC-slichter}. Magnetic phenomena in 2-D materials are of increasing interest, with potential applications in data storage, magnetometry, and quantum information processing \cite{2D_magnetism_review, 2D-Magnetism-review-2, 2D-Magnetism-review-3, 2D-Magnetism-review-4}. However, inductively-detected NMR spectroscopy lacks the sensitivity to probe nanoscale systems \cite{laceyReview}. The nitrogen-vacancy (NV) defect in diamond has emerged as a highly-sensitive probe of magnetism in a variety of materials, with its main advantages being a few-nm sample-sensor distance and a nm$^3$ to $\mu$m$^3$ detection volume \cite{yacoby_review}. Using NVs to directly probe the magnetic and electronic phases of 2-D materials can yield information about magnetic ordering \cite{Gd_pressure,CrIMagnetometry,CrBr3_magnetometry}, fluctuations \cite{demler_2d_sc,demler_magneticIns}, and excitations \cite{Magnon_chem_pot,Magnon_scattering}.

\begin{figure*}[ht]
\centering
\begin{overpic}[width=0.9\textwidth]{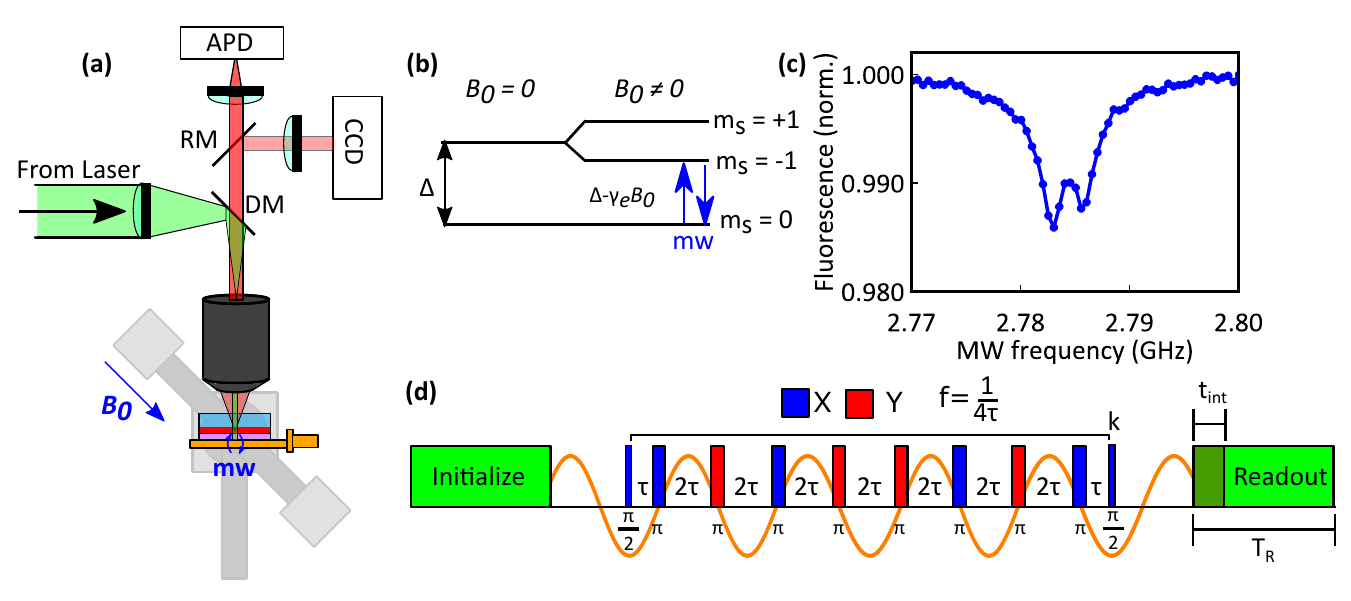}
\end{overpic}
\caption{\label{expSetup} (a) A sketch of the experimental setup. A 532 nm laser excites NVs through an objective. The NV fluorescence is collected through the objective and filtered with a dichroic mirror (DM). The fluoresecence is detected with an avalanche photodiode (APD). The diamond can be imaged optically with a CCD camera and a removable mirror (RM). The NV spin state is controlled with resonant microwave (mw) field with the resonance determined by the bias field $B_0$ directed along the NV axis. (b) The NV center is a S = 1 systems that experiences a crystal field splitting of $\Delta$ = 2.87 GHz splitting the $m_s=0$ and $m_s = \pm1$ sublevels. $B_0$, lifts the degeneracy of the $ m_s = \pm 1$ sublevels. (c) ODMR spectrum of the NV $ m_s = 0 \leftrightarrow -1$ transition. (d) The XY8-N pulse sequence. The spin state is initialized and read out with laser pulses of length $T_R$. The first portion, $t_{\mathrm{int}}$, of the NV fluorescence contains information on the spin state. The 8-pulse block can be repeated $k$ times for a total of $8k=N$ pulses. $\tau$ can be tuned such that the probed frequency matches the Larmor precession frequency of statistically polarized spins.}
\end{figure*}

In one extreme, NV ensembles have been used to detect nuclear spins in $\mu$m$^3$-scale volumes, where the nuclear spin polarization is dominated by Boltzmann statistics due to the bias field and the temperature, the thermal polarization\cite{glenn_high-resolution,smitsNMR}. In the opposite extreme, single NVs are sensitive to nm$^3$-scale volumes, where the nuclear spin polarization is dominated by statistical deviations of the net magnetic moment due to a small spin number, the statistical polarization \cite{statistical_thermal,NVDepthDetermination,varianceNoiseMeriles}. With statistical polarization, the nuclear spins have a small mean magnetic moment, but a substantial magnetic moment variance that can be measured by the NV center.

In this work, we characterize NV ensembles as a function of depth to find the optimal depth to sense statistically-polarized nuclear spins in solids or viscous liquids where diffusion can be ignored. Shallow NVs have the best proximity to the external nuclei, and the NMR signal strength scales as $d^{-3}$, where $d$ is the distance between the NVs and the target nuclei. However, the NV activation rate and spin coherence time improve with increasing depth \cite{NVConversion, depth_decoherence, surface_noise}, suggesting a better magnetic sensitivity for deeper NVs.

We experimentally compare  NV ensembles implanted at different depths to determine the optimal depth for detecting nuclear spins on the diamond surface. We prepared a series of NV ensembles using \nitrogen{} ion implantation with different energies ranging from 1 keV to 7 keV and performed measurements to determine the NV depths and sensitivities to \fluorine{} in Fomblin oil \cite{NVDepthDetermination, devienceMultiSpeciesNMR}. From these measurements, we found that the 5.4 nm deep (2 keV) implant provided the fastest measurement time to obtain a signal-to-noise ratio (SNR) of 3, $ t (\text{SNR} = 3)$ (our figure of merit) when measuring a semi-infinite volume of statistically-polarized \fluorine{} spins.  We then performed NQR spectroscopy on \boron{} nuclei in a hexagonal boron nitride (hBN) flake as a solid-state NV NQR standard and compared the performance of the ensembles to similar NQR detection with single NVs \cite{NVNQRhBN}.

\begin{figure*}
\centering
\begin{overpic}{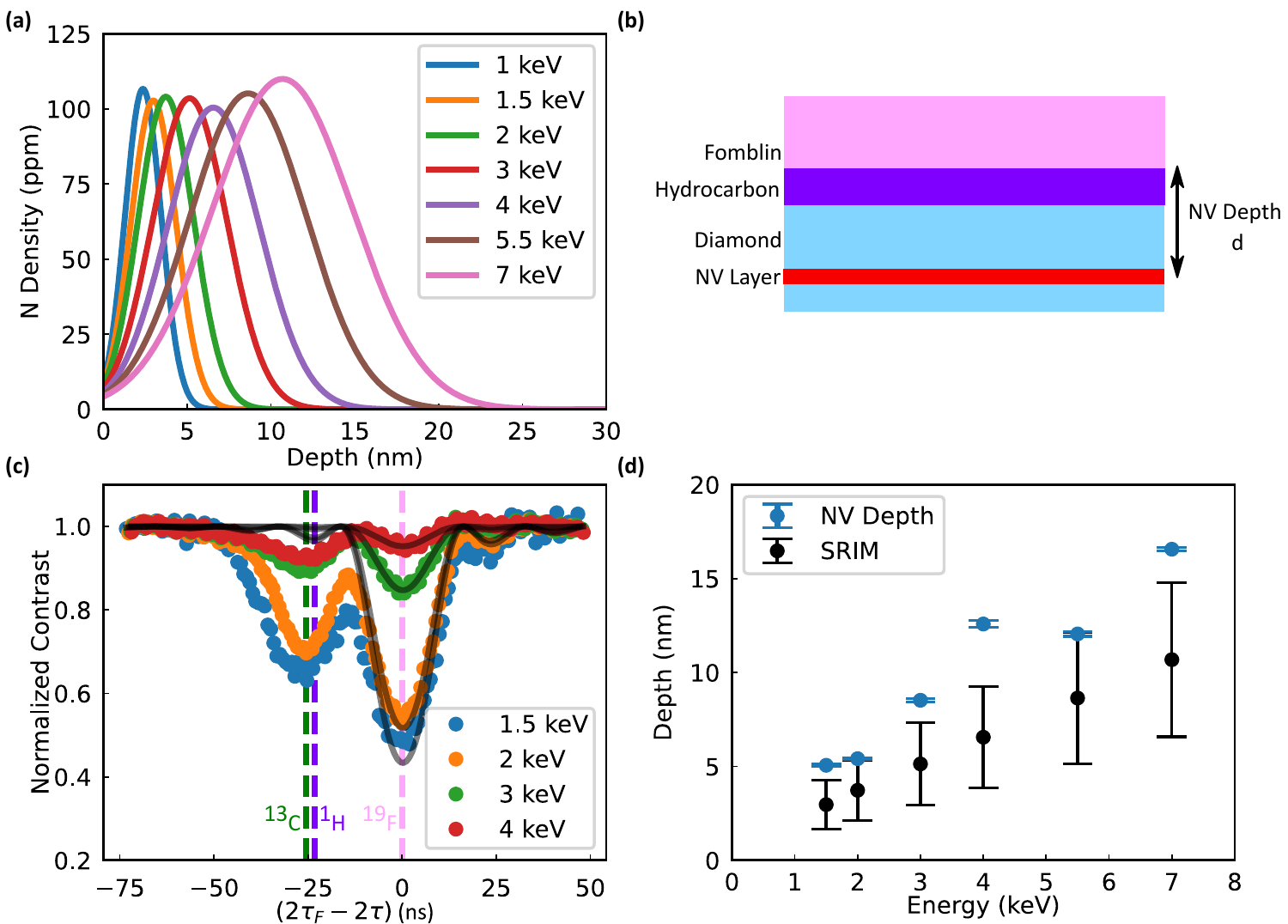}
\end{overpic}
\caption{\label{depthData} (a) A SRIM simulation \cite{SRIMpaper,srimFit} of \nitrogen{} implants in the diamonds used.(b) The sample-diamond interface. The NV layer resides at depth $d$ from the Fomblin. A hydrocarbon layer on the diamond surface separates the NVs from the Fomblin.  (c) The XY8-48 spectra showing $^1$H/$^{13}$C from the surface/diamond and \fluorine{} from Fomblin at different energies. Dashed lines indicate the anticipated 2$\tau$ for \hydrogen{},\fluorine{}, and the $^{13}$C harmonic. Black lines are fits to the \fluorine{} signal using Eqn.~\ref{XY8Contrast}. (d) Determined effective depths using data from (c). Error bars for depths are the standard deviations of a set of five identical measurements. SRIM error bars are the straggle of the simulation.}
\end{figure*}

\section{Experimental Methods}

\subsection{Measurement Techniques}
To characterize our NV ensembles, we use a room-temperature epifluorescence microscope with an NA = 0.8 100$\times$ magnification objective. We excite the NVs with a 40 $\mu$m diameter 532 nm laser beam and NV fluorescence with wavelength longer than 650 nm is detected with a photo-detector (Fig.~\ref{expSetup}(a)). The NV is a spin-1 system with a zero field splitting of $\Delta$ = 2.87 GHz that splits the $m_s = 0$ and $m_s = \pm 1$ sublevels (Fig.~\ref{expSetup}(b)). A pair of magnets provides a bias field $B_0$ aligned to the NV axis to lift the degeneracy of the $m_s = \pm 1$ states. The diamond rests on a copper loop that drives the $m_s = 0 \leftrightarrow -1$ transition with a resonant microwave (mw) field. The NV center exhibits spin-dependent fluorescence rates, fluorescing less in the $m_s = \pm 1$ states than in the $m_s = 0$ state, allowing for optically detected magnetic resonance \cite{NVPhotophysics}. By varying the mw frequency, we observe a decrease in fluorescence at the resonance frequency of the NV $ m_s = 0 \leftrightarrow -1$ transition (Fig.~\ref{expSetup}(c)).

The NV NMR spectroscopy measurements use an XY8-N dynamic decoupling sequence\cite{NVDepthDetermination,protonDD,DDReview} shown in Fig~\ref{expSetup}(d). The NVs are initialized into the $m_s = 0$ state with a  532 nm laser pulse and driven into a superposition with a  mw $\pi$/2 pulse. We apply a train of refocusing $\pi$ pulses preserve the coherence of the spin from noise while amplifying interaction with a narrow frequency range centered at $f = 1/4\tau$ with bandwidth of approximately $\frac{1}{2\tau N}$, where $2\tau$ is the time delay between $\pi$ pulses. The refocusing pulses have phases following the XY8 scheme, $[\pi_x,\pi_y,\pi_x,\pi_y,\pi_y,\pi_x,\pi_y,\pi_x]$, to minimize pulse errors\cite{DDReview}. We repeat this block $k$ times to increase interrogation time of a target frequency, resulting in $8k =N$ total pulses. After the train of $\pi$ pulses, the accumulated phase is mapped on the spin state's populations with a mw $\pi$/2 pulse, then read out with a green laser pulse.

During the XY8-N interrogation time, the NV is sensitive to the magnetic field variance of AC magnetic fields, like that of nuclear Larmor precession, at frequency $f$, directed along the NV axis. For statistically-polarized nuclear spins, each repetition of the sequence measures the Larmor precession with a random initial phase and amplitude. By averaging over many realizations of the nuclear Larmor precession, a dip in NV fluorescence contrast caused by a loss of coherence can be observed when probing the frequency of the nuclei. We use this to measure the magnetic field variance, $B_{\mathrm{RMS}}^2$, of the nuclei and determine the NV depth \cite{protonDD,NVDepthDetermination,degen_QS}.
\begin{figure*}[ht]
\begin{overpic}{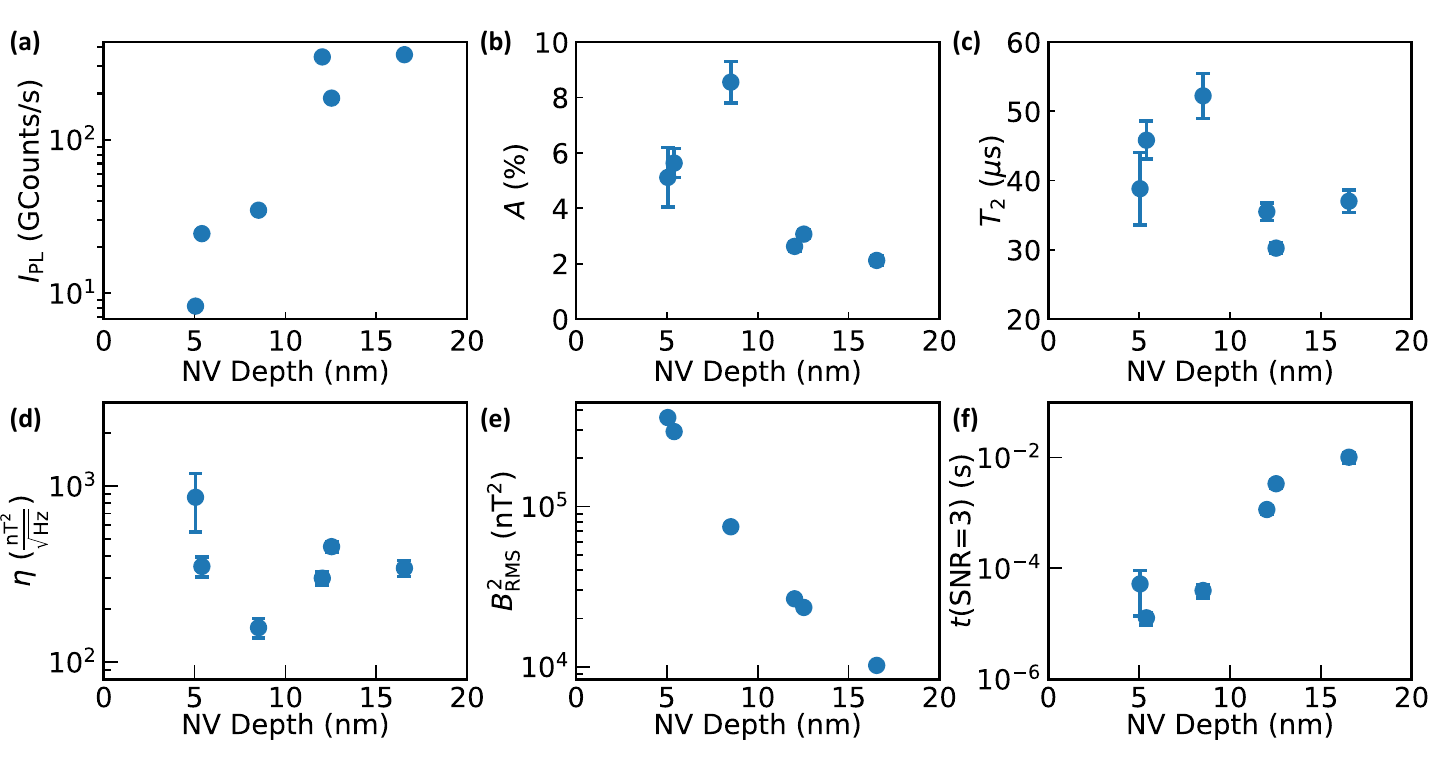}
\end{overpic}
\caption{\label{sensitivity_characterization} (a) PL rate as a function of determined NV depth. (b) The PL spin contrast as a function of determined depth. (c) $T_2$ as a function of determined NV depth for XY8-256. The values in (b) and (c) are the amplitude and time constant, respectively, of an exponential fit of XY8-256 data. Error bars are uncertainties in fit parameters. (d) The sensitivity to magnetic field variance calculated using Eqn.~\ref{sensitivityBrms} as a function of determined NV depth. (e) Magnetic field variance $B_{\mathrm{RMS}}^2$ from \fluorine{} in Fomblin at the measured NV depths. (f) $t$(SNR=3), given the sensitivity in (d), and the $B_{\mathrm{RMS}}^2$ in (e), using Eqn.~\ref{tSNR3}. Error bars may be smaller than markers and unable to be seen.}
\end{figure*}

\begin{table}
\caption{\label{ionEnergyFluence} The energies, fluences, and SRIM-estimated depths for diamonds used in this work. All implants and SRIM simulations are done with an 8\textdegree{} tilt. Fluences are chosen so that the peak concentration is 100 ppm. SRIM depth refers to the mode of the ion distribution.}
\begin{ruledtabular}
\begin{tabular}{lcr}
\hline
Energy (keV) & Fluence (ions/cm$^2$) &SRIM depth (nm)\\
\hline
1.0 & 5$\times10^{12}$ & 2.37\\
1.5 & 6$\times10^{12}$ & 2.97\\
2 & 7.5$\times10^{12}$ & 3.73\\
3 & 1$\times10^{13}$ & 5.13\\
4 & 1.2$\times10^{13}$ & 6.55\\
5.5 & 1.6$\times10^{13}$  & 8.63\\
7 & 2$\times10^{13}$  & 10.68\\
\hline
\end{tabular}
\end{ruledtabular}
\end{table}

\subsection{Diamond Preparation}
We prepared seven 2$\times$2$\times$0.5 mm$^3$ electronic-grade single-crystal diamonds with natural carbon isotope abundance (1.1\% $^{13}$C) from Element Six, which have surface orientation of [100]. Each diamond was implanted with a different energy to provide a  range of NV ensemble depths (see Table~\ref{ionEnergyFluence}). Each fluence was chosen to provide 100 ppm nitrogen concentration at the peak according to Stopping and Range of Ions in Matter (SRIM) calculations (Fig.~\ref{depthData}(a)). We aim for 100 ppm to attain high NV fluorescence intensity while maintaining sufficiently long coherence time for measurements \cite{bauch_NvsT2}. The diamonds undergo a high-vacuum anneal to activate the NV centers and a series of treatments to oxygen-terminate the diamond surface (see supplementary material\cite{Supplemental}). The 1 keV implant had too low fluorescence to do any extensive characterization, which we attributed to low NV concentration.

\section{Results and Discussion}

\subsection{NV depth determination}

For each diamond sample, we applied Fomblin oil to the surface and performed XY8-N measurements. We probed the frequencies surrounding the \fluorine{} Larmor frequency, $\omega_L \approx \gamma_N B_0$, with  $\gamma_N = 2\pi \times$40.08 MHz/T for \fluorine{}, and $B_0 \approx 32$ mT. We determined the depth for each NV ensemble by fitting the fluorescence contrast from an XY8-N measurement as \cite{NVDepthDetermination}: 

\begin{equation}\label{XY8Contrast}
    C(\tau, N) \approx \exp\left(-\frac{2}{\pi^{2}}\gamma_{e}^{2}B^{2}_{\mathrm{RMS}}K(\tau, N)\right),
\end{equation}

\noindent where $\gamma_e$ is the electron gyromagnetic ratio. In this expression, the filter function $K(\tau, N)$, is given by, 

\begin{equation}\label{PhamFilter}
    K(\tau, N) \approx (2\tau N)^2\mathrm{sinc}^2\left[\tau N\left(\omega_L - \frac{\pi}{2\tau}\right)\right],
\end{equation}

\noindent where $\omega_L$ is the \fluorine{} Larmor frequency. Also,

\begin{equation}\label{BrmsFomblin}
    B^{2}_{\mathrm{RMS}}(d) = \rho \left( \frac{\mu_0\hbar\gamma_N}{4\pi}\right)^2 \left(\frac{5\pi}{96d^3}\right)
\end{equation}

\noindent is the magnetic field variance, where $\mu_0$ is the vacuum permeability, $\hbar$ is the reduced Planck's constant, $\rho$ = 4$\times10^{28}$ spins/m$^3$ is the spin density for \fluorine{} in Fomblin oil, and the effective depth $d$ (Fig.~\ref{depthData}(b)) is the distance between the nuclear spins of interest and the NV ensemble. We fit the portion XY8-N spectra associated with \fluorine{} to Eqn.~\ref{XY8Contrast} using $d$ and $\omega_L$ as free parameters (Fig.~\ref{depthData}(c)), with the quantity $d$ displayed in Fig.~2(d). 

We consider $d$ an effective depth because of a hydrocarbon layer on the diamond surface adds more distance between the NVs and the external nuclei \cite{devienceMultiSpeciesNMR}. This hydrocarbon layer can be seen in the XY8-N spectra near the \hydrogen{} resonance, but this peak also corresponds to a harmonic \cite{spurious_harmonics} of the \carbon{} resonance from the 1\% natural abundance of \carbon{} in our diamonds\cite{Supplemental}. Due to the overlap of the \hydrogen{} Larmor frequency and the 4th order harmonic of the \carbon{} frequency, we do not use the hydrogen-related features to determine the NV depths.

The NV depths determined by \fluorine{} were deeper than the SRIM estimates (Fig.~\ref{depthData}(d)). Some physical mechanisms that can result in the sensing NVs being deeper than SRIM estimates include depth-dependent formation efficiency \cite{NVConversion}, the thickness of the hydrocarbon layer\cite{devienceMultiSpeciesNMR}, and band bending converting shallower NVs into the neutral state\cite{band_bending_mapping,dhomkar_charge_2018,charge_instabilities}. Additionally, we are sensing with a distribution of NVs spread over some depth range related to the ion beam straggle. The depth we determine comes from a weighted average of the magnetic field variance over this distribution.  Our findings along with previously observed differences from SRIM estimates\cite{NVDepthDetermination,SpinPropertiesEsnemble} suggest that SRIM simulations cannot be relied upon to accurately determine the depth and must be experimentally determined for any depth-calibrated measurements.

\subsection{Sensitivity Assessment}

In addition to determining the NV depth for each implanted diamond, we measured the continuous wave (cw) photoluminescence (PL) count rate, PL spin contrast, and NV coherence time (Fig.~\ref{sensitivity_characterization}(a)-(c)) to determine the sensitivity as a function of depth. 
We determine NV ensemble sensitivity to a variance field by \cite{nanogratings,NVNQRhBN}:

 \begin{equation}\label{sensitivityBrms}
    \eta \approx \frac{\pi^2e\sqrt{T_2 +T_R}}{\gamma_{e}^2T_2^2A\sqrt{I_{\mathrm{PL}}t_{\mathrm{int}}}},
\end{equation}

\noindent where $I_{\mathrm{PL}}$ is the photon count rate in photons/s, $A$ is the PL spin contrast, $T_2$ is NV coherence time and optimal phase accumulation time, $t_{\mathrm{int}}$ = 2 $\mu$s is the readout signal integration time, and $T_R$ = 10 $\mu$s is the total readout and initialization time.

\begin{figure*}[t]
\begin{overpic}{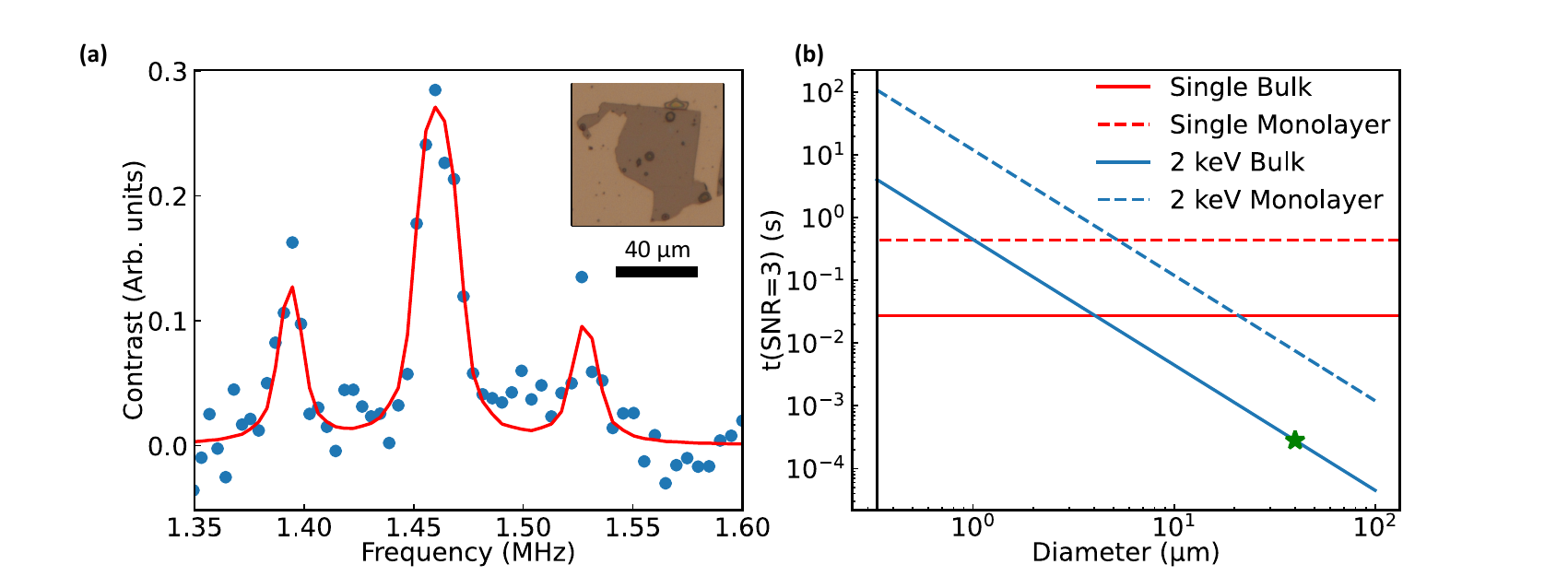}
\end{overpic}
\caption{\label{hbnData} (a)  XY8-256 noise spectroscopy of the \boron{} NQR spectrum from the 5.4 nm deep (2 keV) NV ensemble. The red line is a fit of \boron{} NQR spectrum. The inset is an optical image of a $\sim$100 nm thick hBN flake on the diamond surface. (b) \emph{t}(SNR=3) when detecting bulk and monolayer hBN of a single NV (in red) and the 5.4 nm deep ensemble (in blue) as a function of excitation diameter. The black vertical line on the left is the diffraction limited spot size 532 nm with NA = 0.8. The green star is the \emph{t}(SNR=3) determined for the 5.4 nm deep (2 keV) ensemble.}
\end{figure*}

The photon count rate $I_{\mathrm{PL}}$ (Fig.~\ref{sensitivity_characterization}(a)) is determined by measuring the cw PL intensity on the APD and converting to counts per second. A trend of increasing PL for deeper NV ensembles is observed. We attribute this to the higher nitrogen fluence for deeper implants (see Table~\ref{ionEnergyFluence} and the depth-dependent NV conversion efficiency\cite{NVConversion}.

The contrast and $T_2$ times shown in Fig.~\ref{sensitivity_characterization}(b-c) are from fits with a single exponential of an XY8-256 measurement. The calculated sensitivities are shown in Fig.~\ref{sensitivity_characterization}(d). We find that contrast decreases for deeper NVs, and the coherence times are nearly depth-independent. The 8.5 nm deep NV ensemble exhibits an anomalously good sensitivity due to high contrast and coherence time, with all other NV ensembles having similar sensitivities.

\subsection{Figure of Merit}

With the depth and the sensitivity determined, we use these quantities to calculate our figure of merit, \emph{t}(SNR=3), for a semi-infinite volume of statistically-polarized \fluorine{} (Fig.~\ref{sensitivity_characterization}(d)):
 
 \begin{equation}\label{tSNR3}
    t(\mathrm{SNR}=3) = \left(\frac{3\eta}{B_{\mathrm{RMS}}^2}\right)^2.
\end{equation}
\noindent
The deeper NVs experience a weaker $B^2_{\mathrm{RMS}}$, resulting in a longer \emph{t}(SNR=3) (Fig.~\ref{sensitivity_characterization}(e)). From our data, we find that the shallower NV ensembles offer a much shorter measurement time reaching a minimum with our 5.4 nm deep (2 keV) ensemble. In Fig. \ref{sensitivity_characterization}(d), the most sensitive ensemble has a depth of 8.5 nm. While this is the most sensitive, it does not correspond to the shortest measurement time due to the lower $B_{\mathrm{RMS}}^2$ at that depth.

\subsection{NQR detection of \boron{} in \lowercase{h}BN}

We exfoliated hBN flakes onto the surface of the 5.4 nm (2 keV) diamond with Scotch tape (Fig.~\ref{hbnData}(a)). Any tape residue was cleaned by placing the diamond in an ultraviolet/ozone system for 90 minutes. Before mounting the diamond in the fluorescence microscope, the hBN flake thickness was measured using atomic force microscopy. The flakes measured here were approximately 100 nm thick, effectively infinite for the NV depths we are using.

We detected the NQR spectrum of naturally-abundant \boron{} ($I =3/2$) in hBN using NV centers with a bias field $B_0$ = 2.95 mT with an angle 54.7\textdegree{} relative to the diamond surface normal and the principle axis of the \boron{} electric field gradient. This was done by performing XY8-256 noise spectroscopy near $f = 1/4\tau \approx 1.461$ MHz, the anticipated NQR frequency\cite{NVNQRhBN}. We modeled our NQR spectra (Fig.~\ref{hbnData}(a)) by calculating the transition frequencies and amplitudes of the \boron{} Hamiltonian $\mathcal{H}_{total} = \mathcal{H}_Q + \mathcal{H}_Z$, where $\mathcal{H}_Q = \frac{\nu_Q}{6} (3\hat{I}_z^2 - I(I+1))$ is the uniaxial nuclear quadrupolar Hamiltonian and $\mathcal{H}_Z = \gamma \hat{\vec{I}} \cdot \vec{B}$ is the nuclear Zeeman Hamiltonian. We then determined the resonance frequencies and transition probabilities for each of the nuclear spin transitions. We fit our \boron{} spectrum to this model (see Supplementary Material\cite{Supplemental}) and find $\nu_Q$ = 1.4599 $\pm$ 0.0004 MHz with a $T_2^*$ = 46.9$\pm$ 8.27 $\mu$s (Fig.~\ref{hbnData}(a)), consistent with previous measurements \cite{NVNQRhBN,hBN_NQR_1998,old_hBNNQR}.

\subsection{Ensemble comparison with single-spin detection}

A previously reported experiment of NQR spectroscopy of \boron{} in hBN was performed with single NVs \cite{NVNQRhBN}. Using $T_2$ = 150  $\mu$s, $I_{\mathrm{PL}} = 100$ kCounts/s, $t_{\mathrm{int}} = 250$ ns, $A = 0.35$, and $d = 4$ nm provided in this previous work and others \cite{single_protein_lovchinsky}, we calculate the sensitivity of the single NV and find $\eta \approx 8500 \frac{\mathrm{nT^2}}{\sqrt{\mathrm{Hz}}}$. We calculate the $B^2_{\mathrm{RMS}}$ from \boron{} in an hBN flake for both bulk hBN flakes (100 nm thick) and monolayer hBN (0.4 nm thick). With $B^2_{\mathrm{RMS}}$ and the sensitivity we can calculate the \emph{t}(SNR=3) using Eqn.~\ref{tSNR3} and compare to our optimal NV ensemble. 

The NV ensembles used here offer a sizable improvement over a single NV due to the increase in photon count rate from the large number of NVs in the $D = 40$ $\mu$m diameter excitation region. We also consider $t(\text{SNR}=3)$ as a function $D$ with constant laser power density presuming the only parameter that affects the sensitivity for our NV ensemble is $I_{PL}$. With a uniform distribution of NVs, the photon count rate will vary quadratically with $D$, thus, combining Eqn.~\ref{sensitivityBrms} and~\ref{tSNR3}, $t(\text{SNR}=3) \propto 1/D^2$. We find that our 5.4 nm deep NV ensembles outperforms the single NV for $D$ greater then 4 $\mu$m for bulk, and 5 $\mu$m for the monolayer of hBN (Fig.~\ref{hbnData}).

The 100 ppm implant density of nitrogen in diamond, [N], for this study is not necessarily the most sensitive density for NV-detected NMR spectroscopy. Previous work\cite{bauch_NvsT2} has shown that $1/T_2 \propto B[\text{N}] + 1/T_{2,\mathrm{bg}}$, where $T_2$ is the Hahn echo decoherence time, $B[\text{N}]$ is a relaxation rate per density constant, and $1/T_{2,\mathrm{bg}}$ is a residual background decoherence which is independent of [$\text{N}$]. If $1/T_{2,\mathrm{bg}} \ll B[\text{N}]$, then $T_2 \propto 1/[\text{N}]$. Additionally, the number of NVs, thus, $I_{\mathrm{PL}}$ is also proportional to $[\text{N}]$. This results in $\eta \propto [\text{N}]$. However, if $1/T_{2,\mathrm{bg}} \geq B[\text{N}]$, we find $\eta \propto [\text{N}]^{-1/2}$. This presents an optimum where the sensitivity improvements from $I_{\mathrm{PL}}$ caused by increased $[\text{N}]$ are overtaken by the decrease in coherence time $T_2$. Analysis of this interplay finds a minimum [N] around 0.1-1 ppm depending on the background decoherence rate \cite{Supplemental}.

\section{Conclusions}

We have determined an optimal depth for NMR of statistically polarized semi-infinite volumes assuming standard implantation and annealing procedures. This was accomplished by performing a series of measurements across a set of diamonds with NVs at variable depth. From these measurements, the optimal depth that provides minimal $t$(SNR=3) came from our 5.4 nm deep (2 keV) NV ensemble. We demonstrated the sensing capabilities of these NV ensembles by performing NQR spectroscopy on \boron{} in hBN. We compared our \emph{t}(SNR=3) to that of previous single spin hBN measurements as a function of excitation area and found that NV ensembles outperform single NVs when detecting semi-infinite 2-D materials when $D \geq 4~\mu$m and monolayer materials when $D \geq 5~\mu$m.

\section*{Supplementary Material}
See the Supplementary Material\cite{Supplemental} for details on the experimental setup, diamond processing, optimization of nitrogen density calculations, NMR/NQR spectroscopy with correlation spectroscopy, and fitting of hBN data.
\section*{Acknowledgements}

Sandia National Laboratories is a multi-mission laboratory managed and operated by National Technology and Engineering Solutions of Sandia, LLC, a wholly owned subsidiary of Honeywell International, Inc., for the DOE's National Nuclear Security Administration under contract DE-NA0003525. This work was funded, in part, by the Laboratory Directed Research and Development Program and performed, in part, at the Center for Integrated Nanotechnologies, an Office of Science User Facility operated for the U.S. Department of Energy (DOE) Office of Science. This paper describes objective technical results and analysis. Any subjective views or opinions that might be expressed in the paper do not necessarily represent the views of the U.S. Department of Energy or the United States Government. P.K. is supported by the Sandia National Laboratories Truman Fellowship Program. E.M. was supported by the Department of Defense (DoD) through the National Defense Science \& Engineering Graduate (NDSEG) Fellowship Program. We thank Carlos Meriles for discussions on statistical and thermal polarization and Luca Basso for help with the manuscript.

\section*{Data Availability}
The data that support the findings of this study are available from the corresponding author upon reasonable request.

\bibliography{bib}

\end{document}



\title{Supplementary Material: Nanoscale Solid-State Nuclear Quadrupole Resonance Spectroscopy using Depth-Optimized Nitrogen-Vacancy Ensembles in Diamond}

\author{Jacob Henshaw}\affiliation{\cint}
\author{Pauli Kehayias}\affiliation{\snl}
\author{Maziar Saleh Ziabari}\affiliation{\cint}\affiliation{\unm}
\author{Michael Titze}\affiliation{\snl}
\author{Erin Morissette}\affiliation{\brown}
\author{Kenji Watanabe}\affiliation{\NIMS}
\author{Takashi Taniguchi}\affiliation{\NIMS}
\author{J.I.A. Li}\affiliation{\brown}
\author{Victor M. Acosta}\affiliation{\unm}
\author{Edward Bielejec}\affiliation{\snl}
\author{Michael P. Lilly}\affiliation{\cint}
\author{Andrew M. Mounce}\affiliation{\cint}

\maketitle

\section{Experimental Details}

The  532 nm laser excitation is provided by a 5 W Lighthouse Photonics Sprout laser that is shared across multiple setups. The laser is pulsed using a Gooch \& Housego acousto-optic modulator (AOM). After the AOM, 285 mW of laser power is used for NV illumination. The beam is focused near the back focal plane of a 100$\times$ 0.8 NA objective, resulting in a 40 $\mu$m diameter illumination area on the diamond.  NV fluorescence is collected through the objective and filtered with a 550 nm dichroic mirror and a 650 nm long-pass filter to suppress laser leakage and NV$^0$ fluorescence. The fluorescence is measured using a Thorlabs APD410 avalanche photodiode and digitized with a Keysight M3302A digitizer.

The diamond rests on a copper loop fabricated on a semi-insulating SiC wafer for microwave (mw) spin control. The mw is generated by a Keysight E8267D and the phase is controlled by a Keysight 33600A (250 MSa/s) arbitrary waveform generator (AWG) through IQ modulation. The mw is gated using a switch (Mini-circuits ZASWA-2-50DRA+) and amplified by a Mini-circuits ZHL-16W-43+. For measurements of \boron{} in hBN, a Tabor Proteus (2.5 GSa/s) AWG was used for better timing resolution. The AWG outputs, mw switches, and AOM are triggered by TTL pulses from a PulseBlaster ESR-Pro 500.

The magnetic field is provided by a pair of permanent magnets on a dovetail rail to allow for magnitude control. The dovetail rail is mounted on a pair of rotational stages to align the magnetic field with an NV axis. The field was aligned to the NV axis by measuring the ODMR spectra of the off-axis NVs in the ensemble and overlapping their 3 ODMR peaks.

The pulse sequences used a `quantum interpolation' technique to provide better timing resolution \cite{QuantumInterp}, when using the Keysight 33600A (250 MSa/s) AWG. The last mw pulse of the XY8-N sequence alternates between a $\frac{\pi}{2}$ and a $\frac{3\pi}{2}$ pulse on every readout, then consecutive readouts are subtracted. 

\begin{figure*}[h]
\begin{overpic}{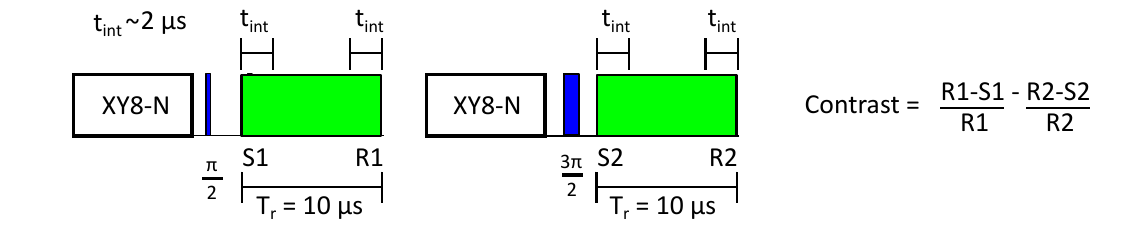}
\end{overpic}
\caption{\label{referencingScheme} The phase cycling and referencing scheme for XY8-N measurements. }
\end{figure*}

Due to the high density of NVs and nitrogen and close proximity to the surface, we use a four-sample readout\cite{wolfSens} method. This is done to suppress any background drift in fluorescence due charge tunneling, as well as suppress any common mode noise from the laser. To do this we perform the phase-cycled readout as described above. Each readout pulse has the first and last $t_{\mathrm{int}}$ of the readout pulse integrated (Fig.~\ref{referencingScheme}). The first of these intervals are considered the signals, S1 and S2, and second last intervals are references, R1 and R2, to the polarized NV ensemble. The signal and the reference of each sequence are subtract and normalized to the reference. The normalized signals of each pair of phase-cycled sequences are subtracted from each other to provide the XY8-N contrast that is analyzed in this work. For for NMR and NQR measurements, the signal is further normalized to a single exponential fit to remove $T_2$ decoherence.

\section{Diamond Preparation Details}

After implantation, the samples were annealed in an ultra-high vacuum furnace ($ < 1\times10^{-8}$ Torr) with the temperature sequence: ramp to 400\textdegree{} C for 2 hours, pause for 2 hours, ramp to 550\textdegree C for 2 hours, pause for 2 hours, ramp to 800\textdegree{} C for 2 hours, pause for 4 hours, ramp to 1100\textdegree{} C for 2 hours, pause for 2 hours, and cooldown to room temperature over 12 hours \cite{SpinPropertiesEsnemble}. The samples were cleaned in a 1:1:1 mixture of sulfuric, nitric, and perchloric acids for 1 hour at 250 \textdegree{}C. Afterwards the samples were place in a ultraviolet/ozone (Uvocs T10X10) system for 90 minutes \cite{UVOzone}. Finally, the samples were annealed in a furnace at 450 \textdegree{}C for 4 hours in an oxygen atmosphere, for oxygen surface termination \cite{SpinPropertiesEsnemble,Sang_dwyer_SurfaceTerm,KMF_selective_oxidation}, and cleaned in a piranha solution (3:1 mixture of H$_2$SO$_4$ and H$_2$O$_2$) heated to 100 \textdegree{}C.

\section{Optimization of Nitrogen density}
As is mentioned in the discussion, the NV sensitivity can be improved by reducing the nitrogen concentration $[N]$. We repeat the equation for sensitivity here to make this clear:
 \begin{equation}\label{sensitivityBrms}
    \eta = \frac{\pi^2e\sqrt{T_2 +T_R}}{\gamma_{e}^2T_2^2A\sqrt{I_{\mathrm{PL}}t_{\mathrm{int}}}}.
\end{equation}
Assuming $T_r \ll  T_2$, the sensitivity to magnetic field variance scales like $T_2^{-\frac{3}{2}}$ with respect to coherence time and $\frac{1}{\sqrt{I_{pl}}}$. More nitrogen can result in more NVs forming, thus higher $I_{pl}$, but shorter $T_2$, potentially resulting in a worse sensitivity. 

Assuming the formation efficiency is independent of nitrogen concentration, we can estimate the sensitivity a sample with lower nitrogen concentration. Using parameters from Bauch \textit{et al.}~\cite{bauch_NvsT2}, we can estimate $T_2$ for Hahn echo measurements as a function of $[N]$. We use the expression
\begin{equation}\label{T2concDependence}
    \frac{1}{T_2} = B_{\textrm{NV-N}}\times[N] + \frac{1}{T_{2,\mathrm{bg}}}.
\end{equation}
In the high $[N]$ regime ($[N]$ \textgreater 1 ppm), $T_2$ varies linearly with $[N]$, with coefficient $B_{\textrm{NV-N}} =$ 160 $\mu$s$\cdot{}$ppm. 

\begin{figure*}[h]
\begin{overpic}{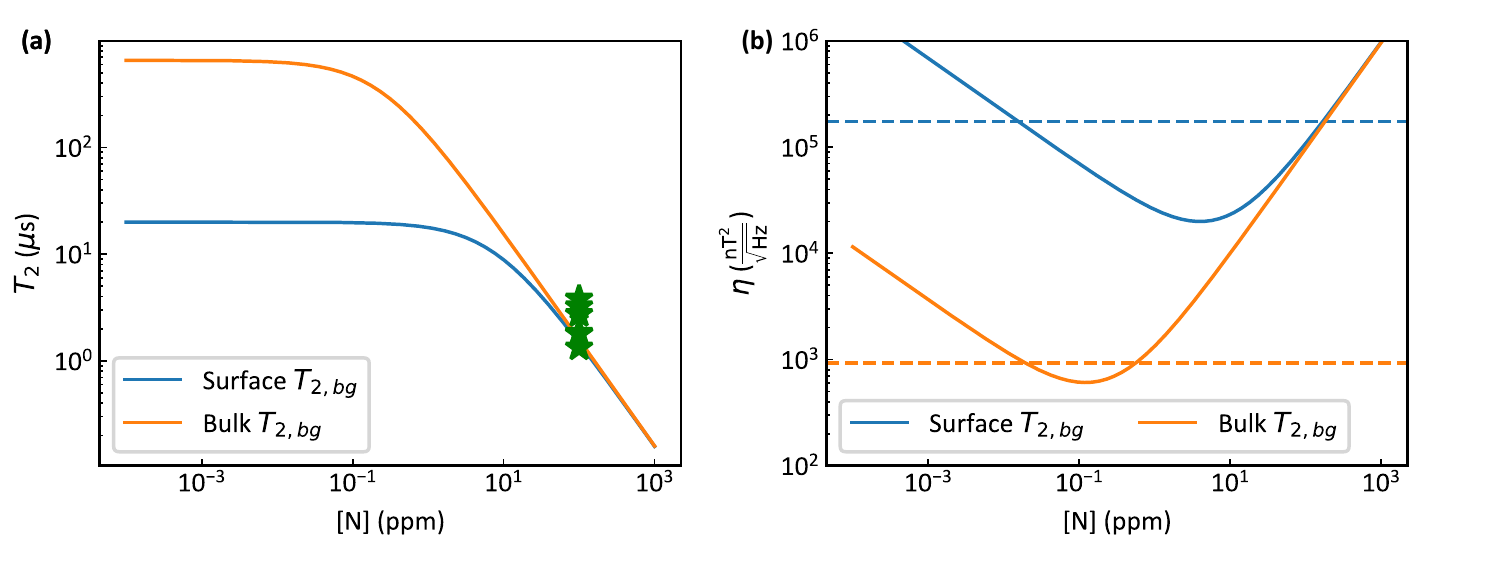}
\end{overpic}
\caption{\label{nitrogenOptimization} (a) Plot of Eqn.~2. $T_{2,\mathrm{bg}}$ for the bulk is 657 $\mu$s and 20 $\mu$s for the surface. Green stars are the Hahn Echo $T_2$'s observed from our samples in this work.  (b) The sensitivity calculated using the $T_2$ times from (a). The dashed lines are the sensitivity of a single NV with the $T_2 = T_{2,\mathrm{bg}}$}
\end{figure*}

The constant $T_{2,\mathrm{bg}}$ was reported to be 657 $\mu$s for NVs in bulk in natural abundance diamond and is attributed to non-nitrogen-based decoherence. A plot of this function is shown in Fig.~\ref{nitrogenOptimization}(a). Myers \textit{et al.}~report a decrease in $T_2$ for depths shallower than 25 nm\cite{surface_noise}, with a $T_2$ of about 20 $\mu$s at around 5 nm. Using the $T_{2,\mathrm{bg}}$ from Myers \textit{et al.}~and Bauch \textit{et al.}~and Eqn.~\ref{T2concDependence}, we can estimate the $T_2$ for different nitrogen concentration, $[N]$ and depth through $T_{2,\mathrm{bg}}$.

With estimates for $T_2$ established, we move to estimate the other relevant parameters in the sensitivity: contrast $A$ and $I_{PL}$. The contrast is estimated to be $A=\frac{0.35}{4e}$. This comes from a single  NV under ideal conditions have roughly 35\% contrast. Then the quantity is divided by 4 to account for contrast reduction due to PL from NVs of different orientation in the ensemble. Finally, the factor of $\frac{1}{e}$, comes from measuring with a phase accumulation time of $T_2$. An estimate for $I_{PL}$ comes from assuming a single NV with a 100$\times$ 0.8 NA 100x magnification objective can expect about 50,000 photons per second from a single NV. We calculate the number of NVs by multiplying $[N]$ by the conversion efficiency, the probability for substitutional nitrogen to become an NV, and the excitation area assuming 40 $\mu$m beam diameter and a thickness of 5 nm defined by the straggle of the ion implantation. This is done by interpolating data from Pezzagna \textit{et al.}\cite{NVConversion}. The NV density is integrated over the NV layer thickness and the beam diameter.

With these estimates, we calculate the sensitivity using Eqn.~\ref{sensitivityBrms} (Fig.~\ref{nitrogenOptimization}(b)). Note that the $T_2$ estimated here is the Hahn echo $T_2$ and can be extended using dynamic decoupling. For the ideal bulk case, the best nitrogen concentration is approximately 0.1 ppm. For the surface background decoherence rate, the most sensitive nitrogen concentration is approximately 5 ppm.

\section{Correlation Spectroscopy Data}

\begin{figure*}[]
\begin{overpic}{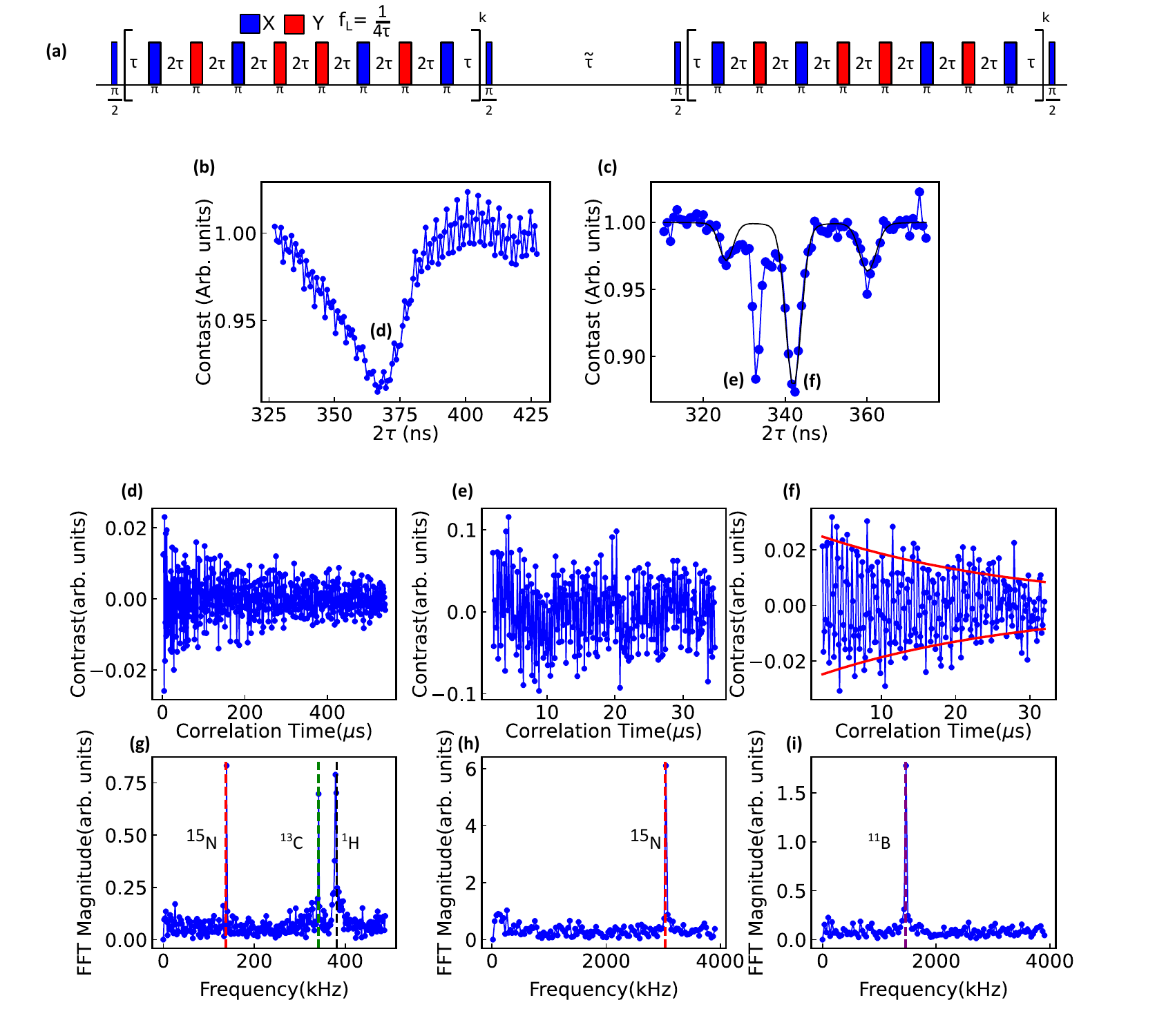}
\end{overpic}
\caption{\label{corrSpec} (a) The correlation spectroscopy pulse sequence. (b) XY8-48 spectrum of the 12.05 nm deep (5.5 keV) sample measuring $^1$H in immersion oil. Fast modulations are artifacts introduced by the quantum interpolation method.(c) Normalized XY8-384 data from the 8.5 nm (3 keV) (d-f) Time domain measurements using correlation spectroscopy of the corresponding contrast dips in (b) and (c).The red lines in (f) are exponential envelopes. With time constant 27.8 $\mu$s. (g-i) Fourier transforms of (d-f) respectively. The dashed line indicate calculated frequencies for various nuclei given the magnetic field and sampling rate with correlation spectroscopy.}
\end{figure*}

In the main text we point out a few shortcomings of the XY8-N noise spectroscopy technique, specifically harmonics showing signals at frequencies other than the Larmor frequency of a nuclei. One method to determine the origin of a signal is correlation spectroscopy\cite{spurious_harmonics}. We implement correlation spectroscopy by performing two XY8-N pulse sequences in a row, with the timing delay fixed to half the Larmor period of a target nuclei or somewhere where a dip in contrast is observed in normal XY8-N spectroscopy, and varying the time delay between the two sequences (Fig.~\ref{corrSpec}(a)). Correlation spectroscopy can provide the equivalent of a Free Induction Decay (FID) from inductively-detected NMR\cite{13C_correlation,Proton_correlation,nanogratings}. This is both very powerful and a limitation.

For nuclei with $T_2^* \gg T_{1,\mathrm{NV}}$, like $^{19}$F or $^1$H, this method can provide $1/T_{1,\mathrm{NV}}$-limited frequency resolution, typically on the order of ~500 Hz. However,  the  correlation spectroscopy signal is reminiscent to an FID, the nuclear spin dephasing time can also be a limiting factor. This makes correlation spectroscopy a useful (though not particularly sensitive) tool for measuring nuclear dephasing times. Since the delay time $\widetilde{\tau}$ needs to be extended in order to gain spectral resolution, this approach trades sensitivity for spectral resolution. There are other implementations of correlation spectroscopy, such as continuous sampling \cite{glenn_high-resolution}, which can be implemented without a significant sacrifice in sensitivity, but lose on spectral bandwidth.

Correlation spectroscopy provides an unambiguous way to identify the origin of decoherence dips observed in XY8-N spectroscopy. This is best shown when looking for $^1$H in immersion oil using correlation spectroscopy. The XY8-48 data shown in Fig.~\ref{corrSpec}(b) displays an asymmetric line that, in principle, should be caused by $^{1}$H. Correlation spectroscopy reveals a few interesting features. In Fig.~\ref{corrSpec}(d) an exponential enveloped signal can be observed. More detail can be extracted by Fourier transforming the signal (Fig.~\ref{corrSpec}(g)). The Fourier transform revels two additional frequencies, one signal at the $^{15}$N Larmor frequency, another at the $^{13}$C Larmor frequency, and a broad line at the folded Larmor frequency of $^1$H. The width of this line can be attributed to molecular diffusion in the immersion oil which limits the $T_2^*$\cite{Proton_correlation}. These additional contributions are the reason why we do not use immersion oil or surface hydrogen to determine the depth.

In the main text, we point out an additional feature in our hBN spectra (Fig.~\ref{corrSpec}(c)). Correlation spectroscopy of this feature and Fourier transforming the time trace reveals a line at $\sim$3 MHz (Fig.~\ref{corrSpec}(e,h)). This originates from the resonant frequency of the NV's host $^{15}$N when the NV is in the $m_0=\pm1$, enabling a 3.05 MHz hyperfine coupling. We also perform correlation spectroscopy on the another major feature in the hBN spectra (Fig.~\ref{corrSpec}(f,i)). We are able to estimate the $T_2^*$ of $^{11}$B from the decay envelope of the time trace. This is in good agreement with the XY8-384 line width.

\section{Fitting of \boron{} in \lowercase{h}BN data}
To fit the data in Fig.~4(a) in the main text and Fig.~\ref{corrSpec}(c) here, we use Eqn.~1 of the main text. However, since \boron{} has a $T_2^*$ that is on the order of the NV $T_2$ and also is a spin-3/2 system ($I=3/2$), the filter function $K(\tau,N)$ and $B_{\textrm{RMS}}^2$ must be generalized. The filter function, Eqn.~2 in the main text, assumes that $T_{2,n}^* \gg 2\tau N$. To account for this we use a modified filter function, $K(\tau, N, T_2^*)$, derived in Ref.~\cite{NVDepthDetermination}:

\begin{equation}
\begin{split}
    K(\tau, N,T_{2,n}^*) &\approx \frac{2T_{2,n}^{*2}}{\left[1 + T_{2,n}^{*2}\left(\omega_L-\frac{\pi}{2 \tau}\right)^2\right]^2}\\
                           &\times \Big( e^{-\frac{2\tau N}{T_{2,n}^*}} \Big\{ \left[1-T_{2,n}^{*2}\left(\omega_L-\frac{\pi}{2 \tau}\right)^2\right]\\
                           &\times \text{cos}\left[2\tau N\left(\omega_L -\frac{\pi}{2 \tau}\right)\right]\\
                           &-2T_{2,n}^{*2}\left(\omega_L-\frac{\pi}{2 \tau}\right)\text{sin}\left[2\tau N\left(\omega_L-\frac{\pi}{2 \tau}\right)\right] \Big\}\\
                           &+ \frac{2\tau N}{T_{2,n}^{*}}\left[1 + T_{2,n}^{*2}\left(\omega_L-\frac{\pi}{2 \tau}\right)^2\right]\\
                           &+T_{2,n}^{*2}\left(\omega_L-\frac{\pi}{2 \tau}\right)^2-1\Big) ~.
\end{split}
\end{equation}

Due to the additional complexity of the \boron{} Hamiltonian, the evaluation of the $B_{\textrm{RMS}}^2$ is more involved. We start by finding the eigenvectors and eigenvalues of the \boron{} Hamiltonian
\begin{equation}\label{Boron_Hamiltonian}
    \mathcal{H}_{total} = \mathcal{H}_Q + \mathcal{H}_Z = \frac{\nu_Q}{6} (3\hat{I}_z^2 - I(I+1)) + \gamma \hat{I} \cdot \vec{B}.
\end{equation}
It is important to note that in the frame of the \boron{}, the magnetic field is ideally at a $\sim$54.7\textdegree{} angle relative to the principle axis of the electric field gradient (EFG) of the \boron{} nucleus. 

To calculate the $B_{\textrm{RMS}}^2$, we repeat the process laid out in Ref.~\cite{NVDepthDetermination} and the supplemental material of Ref.~\cite{NVNQRhBN}. The $B_{\textrm{RMS}}^2$ is given by
\begin{equation}\label{hBN_Brms}
    B_{\textrm{RMS}}^2 = 9\rho \left(\frac{\mu_B\hbar\gamma_n}{4\pi}\right)^2\left(\Gamma_x(\alpha) f^{x,x}(\omega) + \Gamma_y(\alpha) f^{y,y}(\omega)\right) .
\end{equation}
In this expression, $\Gamma_{x,y}$ are geometric factors depending on the angle of the magnetic field $\alpha$, the NV depth $d$, and the hBN thickness, h. These are given by
\begin{equation}\label{gammaX}
    \Gamma_x(\alpha) = -\frac{\pi}{3}\left(\frac{3\mathrm{cos}(4\alpha)-35}{768}\right)\left[\frac{1}{d^3}-\frac{1}{(d+h)^3}\right] ,
\end{equation} 
\begin{equation}\label{gammaY}
    \Gamma_y(\alpha) = \frac{\pi}{3}\left(\frac{3\mathrm{cos}(2\alpha)+5}{192}\right)\left[\frac{1}{d^3}-\frac{1}{(d+h)^3}\right] .
\end{equation}
The quantities $f^{x,x}$ and $f^{y,y}$ are the spin-spin correlation functions given by the following:
\begin{equation}\label{spin_correlators}
    f^{\alpha,\alpha}(\omega) = \frac{2}{\mathrm{Tr} (\textbf{1})}\sum_{n,m}|\bra{n_z}I_\alpha\ket{m_z}|^2\delta(\omega_L-\omega) .
\end{equation}
Here $\alpha = x,y$,  $I_\alpha$ are the $x$ or $y$ spin operators for a spin $I$ system. For \boron{}, $I$ = 3/2, $\ket{n_z}$ is the eigenvector with eigenvalue $n_z$.

With the $B_{\textrm{RMS}}^2$ and $K(\tau, N, T_2^*)$ defined, we fix $h$ at 100 nm, and leave $d$, $\vec{B}$, and $T_2^*$ as free parameters and fit the data to to the contrast function in the main text (Eqn.~1).
\bibliography{bib}